\documentstyle[psfig,12pt,aasms4]{article}
\begin{document}
\title{Discovery of a $\sim$1 Hz quasi-periodic oscillation in
the low-mass X-ray binary 4U\,1746--37}

\author{Peter G. Jonker\altaffilmark{1}, Michiel van der
Klis\altaffilmark{1}, Jeroen Homan\altaffilmark{1}, Rudy
Wijnands\altaffilmark{1,2}, Jan van
Paradijs\altaffilmark{1}$^{,}$\altaffilmark{3}, Mariano
M\'endez\altaffilmark{1,4}, Erik Kuulkers\altaffilmark{5,6}, Eric
C. Ford\altaffilmark{1}} \altaffiltext{1}{Astronomical Institute
``Anton Pannekoek'', University of Amsterdam, and Center for
High-Energy~Astrophysics,~Kruislaan 403, 1098 SJ Amsterdam;
peterj@astro.uva.nl, michiel@astro.uva.nl, homan@astro.uva.nl,
rudy@astro.uva.nl, jvp@astro.uva.nl, mariano@astro.uva.nl,
ecford@astro.uva.nl} \altaffiltext{2}{MIT, Center for Space Research,
Cambridge, MA 02139} \altaffiltext{3}{University of Alabama,
Huntsville} \altaffiltext{4}{Facultad de Ciencias Astron\'omicas y
Geof\'{\i}sicas, Universidad Nacional de La Plata, Paseo del Bosque
S/N, 1900 La Plata, Argentina} \altaffiltext{5}{Space Research
Organization Netherlands, Sorbonnelaan 2, 3584 CA Utrecht, The
Netherlands; e.kuulkers@sron.nl} \altaffiltext{6}{Astronomical
Institute, Utrecht University, P.O. Box 80000, 3507 TA Utrecht, The
Netherlands}

\subjectheadings{accretion, accretion disks --- stars: individual
(4U~1746--37) --- stars: neutron --- X-rays: stars}
\begin{abstract}
We have discovered a $\sim$1 Hz quasi-periodic oscillation (QPO) in
the persistent X-ray emission and during type I X-ray bursts of the
globular cluster source, dipper and low-mass X-ray binary (LMXB)
4U~1746--37. The QPO properties resemble those of QPOs found recently
in the LMXB dippers 4U~1323--62, and EXO~0748--676, which makes
4U~1746--37 the third source known to exhibit this type of QPOs.  We
present evidence for X-ray spectral changes in this source similar to
those observed in LMXBs referred to as atoll sources. We detect two
states, a low intensity and spectrally hard state, and a higher
intensity and spectrally soft state.  This may explain the different
spectral characteristics reported for 4U 1746-37 earlier. The high
intensity state resembles the banana branch state of atoll
sources. The QPOs are only seen in the low intensity state, and are
absent when the source is in the banana branch. This strongly suggests
that either the accretion disk or an extended central source change
shape between the low intensity state and the banana branch. \par

Twelve bursts were detected, of which 5 took place while the source
was on the banana branch and 7 when the source was in the low
intensity state. The bursts occurring on the banana branch had an
e-folding time $\sim$3 times longer than those which occurred in the
low intensity state. \par

Whereas previously detected dips showed only a decrease in count rate of
$\sim$15\%, we found in one observation a dip in which the count rate
dropped from $\sim$200 counts per second to $\sim$20 counts per
second. This dip lasted only $\sim$250 seconds, during which clear
spectral hardening occured. This is the first time strong evidence for
spectral changes during a dip are reported for this source.\par
\end{abstract}

\section{Introduction}
The low-mass X-ray binary (LMXB) 4U~1746--37 shows type I X-ray bursts
(Li \& Clark 1977) and dips (Parmar et al. 1989) which recur on the
5.7 hours (Sansom et al. 1993) orbital period.  The source is located
in the globular cluster NGC~6441. Deutsch et al. (1998) reported the
probable detection of the optical counterpart. The dips in dip sources
such as 4U~1746--37 are thought to be caused by periodic obscuration
of the central source by a structure formed in an interplay between
the accretion stream from the companion towards the disk and the disk
itself (White \& Swank 1982; White, Nagase, \& Parmar 1995; Frank,
King, \& Lasota 1987). This structure moves through the line of sight
fully or partially obscuring the central source once per binary
cycle. Often, the X-ray spectral properties change during the
dips. Spectral hardening occurs in 4U~1323--62 (Parmar et al. 1989),
and spectral softening in deep dips seen in 4U~1624--49 (Church \&
Baluci\'nska-Church 1995). In 4U~1746--37 no energy dependence was
found so far (Sansom et al. 1993).  The type I X-ray bursts are
observed to have peak luminosities around the Eddington luminosity for
the estimated distance to the globular cluster (10.7 kpc; Djorgovski
1993). Combined with the fact that the ratio of the X-ray to optical
luminosities $\rm{L_x/L_{opt} \sim 10^3}$, this makes it likely that
outside the dips the central source is viewed directly (Parmar et
al. 1999). \par

Recently, in two dipping LMXBs, 4U~1323--62 (Jonker et al. 1999a) and
EXO~0748--676 (Homan et al. 1999), quasi-periodic oscillations (QPOs)
with a typical frequency of 1 Hz were discovered. In EXO~0748--676
these frequencies vary between 0.58--2.44 Hz on timescales ranging
from several days to weeks (Homan et al. 1999). Observations with
intervals of several days to weeks revealed similar frequency shifts
in 4U~1323--62 (Jonker et al. 1999b).  These QPOs are strong, with
fractional rms amplitudes of $\sim$10\%, which are the same in the
persistent emission, during dips, and during type I X-ray bursts. The
rms amplitude of the QPOs does not, or only weakly, depend on the
photon energy, unlike that of various other types of QPO observed from
LMXBs (van der Klis 1995, 1998).  From these properties, and the fact
that the $\sim$1 Hz QPOs are observed in high inclination systems,
Jonker et al. (1999a) suggested that partial obscuration of the
central source by a near-opaque or gray medium in or on the disk
covering and uncovering the source quasi-periodically at the observed
QPO frequency could explain the QPO properties. The most natural
location for such a medium would be at the radius where the orbital
frequency in the disk is equal to the QPO frequency, which corresponds
to a radius of $\sim2$~x~$10^8$cm for a 1.4$M_{\odot}$ neutron star
(Jonker et al. 1999a). \par

On the basis of the timing and spectral properties of LMXBs Hasinger
\& van der Klis (1989) defined the atoll and the Z sources. The atoll
sources show at least two states, named after the associated
structures in the X-ray color-color diagram (CD): the island and the
banana branch. The timing and spectral properties of the sources in
these states are different. In the island state strong band limited
noise is present in the power spectra. When the source moves up the
banana branch the fractional rms amplitude of this noise component
decreases and simultaneously the rms amplitude of a very low frequency
power law noise (VLFN) component increases. The Z sources have higher
luminosities and trace out a Z track in the CD. The timing properties
are correlated with the position along this track (for a complete
overview of the timing and spectral properties of the atoll and Z
sources, see van der Klis 1995).\par

In this Letter, we report the discovery of a $\sim$1 Hz QPO in the
source 4U~1746--37. We also show that the spectral characteristics and
some of the timing properties of this source are reminiscent of those
of an atoll source.

\section{Observations and analysis}
\label{analyse}
We analysed 9 observations of 4U~1746--37 obtained with the
proportional counter array (PCA) on board the {\em RXTE} satellite
(Bradt, Rothschild, \& Swank 1993), in 1996 on October 25, 27, 31, and
in 1998 on June 5, June 6, August 3, November 7, and November 22. The
total amount of good data analysed was $\sim$129 ksec. A log of the
observations can be found in Table~\ref{obs_log}. Data were obtained
over an energy range of 2--60 keV simultaneously with time resolutions
of 0.125~s, 16~s, and 1~$\mu$s in 1, 129, and 255 energy bins; the
Standard 1, Standard 2, and GoodXenon data modes, respectively.  \par

\begin{deluxetable}{cccccc}
\tablecaption{Log of the observations. The 5 detector average 2--60
keV persistent emission count rate of each observation is given in
column 6.\label{obs_log}}

\startdata
Number & Observation & Date & Start time & Amount of &Persistent \nl
       &   ID        & &   & good data  & emission  \nl
       &             & &  (UTC) & (ksec.) & (c/s) \nl
\tableline
1 &10112-01-01-00 & October 25 1996& 00:10:55&$\sim$22 & 73 \nl
2 &10112-01-01-01 & October 27 1996& 03:25:07&$\sim$17 & 73 \nl
3 &10112-01-01-02 & October 31 1996& 08:26:19&$\sim$21 & 66 \nl
4 &30701-11-01-00 & June 5   1998& 00:09:55& $\sim$6 & 365 \nl
5 &30701-11-01-01 & June 5   1998& 14:09:03& $\sim$15 & 415 \nl
6 &30701-11-01-02 & June 6   1998& 07:45:43& $\sim$3 & 414 \nl
7 &30701-11-02-00 & August 3 1998& 03:15:03& $\sim$25 & 296 \nl
8 &30701-11-03-01 & November 7 1998& 08:19:37& $\sim$7 & 728 \nl
9 &30701-11-04-00 & November 22 1998& 11:18:33& $\sim$13 & 785 \nl
\enddata
\end{deluxetable}

\begin{figure}[bh]
\centerline{\psfig{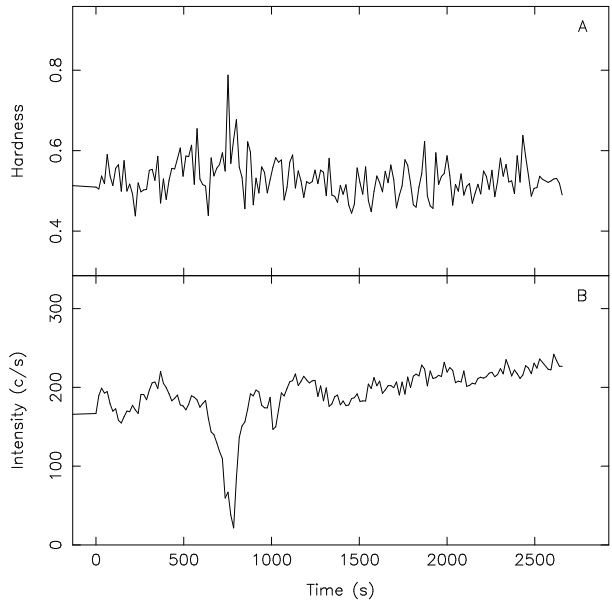}}
\figcaption{\label{dip}
A: Hardness curve (ratio between the count rates in the 9.7--16 keV
and the 2--9.7 keV bands) of the last part of observation 7 (time zero
corresponds to August 3 14:15:03 UTC). B: The simultaneous 2--60 keV,
background-subtracted light curve. There is clear spectral hardening
during the dip apparent in the lower panel. No dead time corrections
were applied (the deadtime fraction was $<$1.5\%). Since the
photon-photon deadtime is completely negligible both in and outside
the dip, the deadtime does not affect the hardening during the dip.}
\end{figure}

The average persistent emission count rates varied from 66 counts per
second (observation 3) to nearly 800 counts per second (observation
9; see Table~\ref{obs_log}). All count rates reported are background
subtracted and for five PCA detectors, unless otherwise stated. We
observed shallow dips in the lightcurve showing a decrease in count
rate of only $\sim$15\%, and one deep dip with evidence for spectral
hardening (see Fig.~\ref{dip}, and Section 3).  During the last part
of observation 9 a solar flare occured; we excluded these data from our
analysis. \par

Since the dips were often difficult to distinguish from the persistent
emission (Sansom et al. 1993; Parmar et al. 1999), we divided the
data into just two categories; non-burst (which combines persistent
emission and dip data) and burst data.  \par

For the selection of the burst data the start of a burst was
characterized by the sharp increase in flux and simultaneous decrease
in hardness. The burst e-folding time was determined by fitting an
exponential to the burst decay in the 2--60 keV lightcurve of the
Standard 1 data. The end of the burst was chosen to be three times the
e-folding time after the onset (see also Jonker et al. 1999a). \par

Using the Good Xenon data, we calculated power spectra of non-burst
data segments of 64~s length and of burst data segments of 16~s length
in the energy bands 2--60 keV, 2--6.4 keV, and 6.4--60 keV, with a
Nyquist frequency of 512 Hz. We also calculated power spectra of
segments of 16~s length with a Nyquist frequency of 2048 Hz in the
2--60 keV band. The power spectra within each energy band were added
and averaged for each observation, separately for burst and non-burst
data. To search for burst oscillations we also calculated power
spectra of a length of 1~s with a Nyquist frequency of 2048 Hz. \par

We constructed a CD from all non-burst Standard 2 data, using four
detectors (0, 1, 2, and 3). The hard color in this diagram is defined
as the logarithm of the ratio between the count rates in the 9.7--16.0
keV band and the count rates in the 6.0--9.7 keV band, and the soft
color as the logarithm of the ratio between the count rates in the
3.5--6.0 keV band and the count rates in the 2--3.5 keV band \par

\section{Results}
\begin{figure}[bh]
\centerline{\psfig{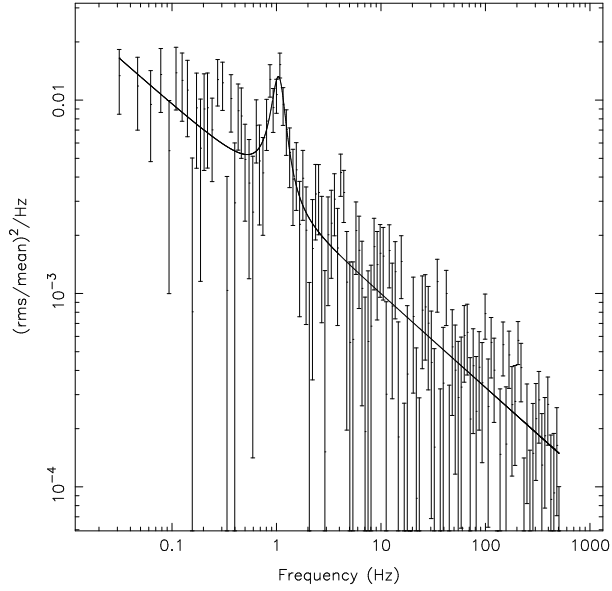}}
\figcaption{Normalized (van der Klis 1988) power spectrum of
observation 1. The Poisson noise has been subtracted. The solid line
represents the best fit to the data using two components in the fit: a
Lorentzian at $\sim$1 Hz (the QPO), and a power law.
\label{qpo}}
\end{figure}

In the non-burst data of October 25 1996 (observation 1), we
discovered a QPO in the 2--60 keV band with a frequency, FWHM, and rms
amplitude of $1.04\pm0.03$ Hz, $0.36\pm0.09$ Hz, and 7.7\%$\pm$0.7\%
(6$\sigma$), respectively (see Fig.~\ref{qpo}). In the observations
of October 27 (observation 2) and October 31 (observation 3) 1996, we
observed QPOs with similar frequencies, FWHMs and rms amplitudes,
albeit at a lower significance level (see Table~\ref{sig_det}). \par

\begin{deluxetable}{cccccc}
\tablecaption{The fractional rms amplitudes, frequencies and FWHMs of
the QPO in the observations where 4U~1746--37 was observed in the low
intensity state. If no QPO at a significance level higher than
3$\sigma$ was detected, the 95\% confidence upper limit is given. The
FWHM and the frequency were measured in the 2--60 keV band, except for
observation 2, where they were measured in the 2--6.4 keV
band. \label{sig_det}}

\startdata

Observation & rms amplitude & rms amplitude & rms amplitude &
$\nu_{QPO}$ & FWHM \nl
     & (2--60 keV) & (2--6.4 keV) & (6.4--20 keV) & (Hz) & (Hz) \nl
\tableline
1& 7.7$\%\pm0.7$\% & 7.8\%$\pm0.9$\% &
6.7$\%^{+2.0}_{-1.1}$\% & $1.04\pm0.03$ & $0.36\pm0.09$ \nl
2& $<7$\% & 6.9\%$^{+1.2}_{-0.9}$\% & $<$6.8\% & 1.59$\pm0.05$
 & 0.4$\pm0.2$ \nl 
3& 7.5\%$^{+1.4}_{-1.1}$\% & $<$7.4\% &$<$8.7\% &
1.01$\pm0.06$ & 0.25$\pm0.13$ \nl
\enddata
\end{deluxetable}

We measured the QPO properties in the 2--6.4 keV and 6.4--60 keV
bands. Upper limits on the presence of the QPO were derived fixing the
values of the FWHM and frequency obtained in the 2--60 keV energy
band, except for observation 2. For this observation we used the
values obtained in the 2--6.4 keV band, since the QPO was not detected
at a significance higher than $3\sigma$ in the 2--60 keV band (see
Table~\ref{sig_det}). Only for observation 1 the QPO was detected at a
significance higher than 3$\sigma$ in both the 2--6.4 keV and the
6.4--20 keV energy band. Within the errors the rms amplitude of the
QPO in the two energy bands was the same (see
Table~\ref{sig_det}). \par

The QPO was also detected when all 16~s segment power spectra (9 in
total) of the decay of the three bursts of observation 1 were
combined, at a similar fractional rms amplitude (7.2\%$\pm0.8$\%). The
averaged count rate was $\sim$720 counts per second, i.e., about ten
times that in the persistent emission. The frequency increased
significantly with respect to the frequency of the QPO in the
persistent emission to $1.39^{+0.05}_{-0.10}$ Hz. The FWHM of the QPO
was consistent with being the same as in the persistent emission. In
the bursts of observations 2 and 3 no significant QPO was detected,
with upper limits on the rms amplitude of 7\% to 8\%. \par

We fitted a power law to the noise component evident in
Fig.~\ref{qpo}. Its rms amplitude integrated over 0.0156--1 Hz varied
from 7.1\%$\pm0.3$\% to 5.9\%$\pm0.4$\% and to 9.3\%$\pm0.4$\%, from
observation 1--3, while the power law index was consistent with being
constant at $0.5\pm0.3$. In the observations 4--9, the noise amplitude
was in the range of 1.6\%--5.7\%. Here, the fractional rms amplitude
increased with increasing intensities (see Table~\ref{obs_log}),
except for observation 8 were it was only 1.6\%. The power law indices
ranged from $1.7\pm0.4$ to $2.1\pm0.4$. The rms amplitude of this
power law component is slightly higher in the 6.4--20 than in the
2--6.4 keV band. The 6.4--20/2--6.4 keV spectral hardness did not 
significantly depend on the presence or absence of the QPO.\par

\begin{figure}[bh]
\centerline{\psfig{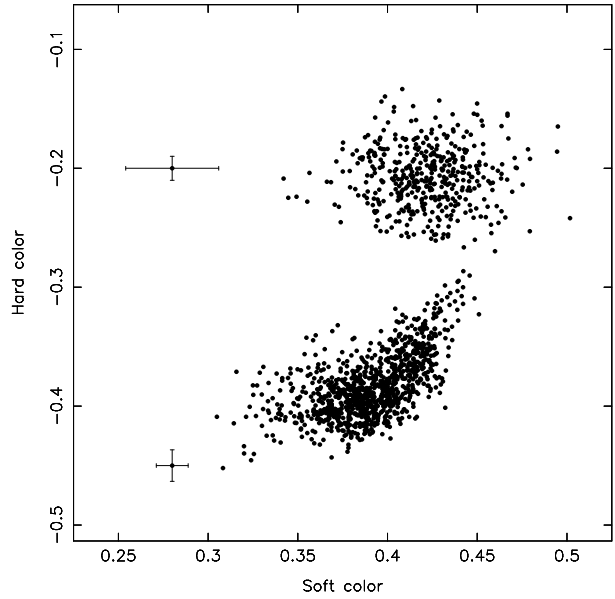}}
\figcaption{Color-color diagram of 4U~1746--37. The hard color is
defined as the log of the 9.7--16.0 keV / 6.0--9.7 keV count rate
ratio, the soft color as the log of the 3.5--6.0 keV / 2--3.5 keV
ratio. No dead-time corrections have been applied (the deadtime
fraction was $<$2\%). The data are background subtracted and bursts
were removed. The data points are 64~s and 128~s averages for the
banana branch (lower part) and for the presumed (see text) island
state (upper part), respectively. Typical error bars are shown for
data in the hard (island) state and for data on the banana
branch.\label{cd}}
\end{figure}

Combining all non-burst data into a CD, we observed a pattern which
resembles the atoll shape (Hasinger \& van der Klis 1989),
(Fig.~\ref{cd}). Two distinct parts can be distinguished in the CD, a
hard (low intensity, upper part of the diagram) island-like state, and
a softer (higher intensity, lower part of the diagram) banana branch.
The $\sim$1 Hz oscillations were only found when the source was in the
upper (hard) part of the diagram (made up of observations 1--3). Upper
limits on the presence of a QPO in the range of 0.5--2.5 Hz of
1\%--2\% were derived for the banana part of the diagram (made up of
observations 4--9).  \par

We observed a total of 12 type I bursts. One burst in the second
observation and two bursts in the third observation showed secondary
bursts, occurring several hunderd seconds after the first burst and
with a 3--4 times lower peak flux (see Lewin, van Paradijs, \& Taam
1995). The burst e-folding times ranged from 7.5 to 11.3 seconds for
bursts observed in observations 1--3, and from 7.5 to 14.5 for the
secondary bursts. The burst e-folding time of 4 of the 5 bursts
observed in observation 8 and 9 ranged from 27.7--30.1 seconds, while
one burst in observation 9 which occured $\sim$600 seconds after a
burst with a 2.3 higher peak flux had an e-folding time of
only 3.0 seconds (the peak flux of this very short burst was similar
to that of the bursts in the low intensity state). Upper limits on
burst oscillations in the 100--1000 Hz frequency range during the peak
and decay of the bursts adding the different bursts of typically
$\sim$15\% were derived. \par

One of the dips observed in observation 7 was different from the
others; it showed a simultaneous sharp increase in hard color of
$\sim$28\%, and drop in count rate from $\sim$200 counts per second to
$\sim$21 counts per second. The dip was quite short ($\sim$250 s),
whereas previous dips were shallow ($\sim$15\%), and exhibited no clear
spectral hardening (see Fig.~\ref{dip}). Due to the uncertainties in
the orbital ephemeris, we could not relate the exact time of this deep
dip to the dips in the other observations. \par

Upper limits on the presence of kHz QPOs in the frequency range of
100--1000 Hz were derived using a fixed FWHM of 50 Hz. The upper
limits were in the range of 3.4\%--6.5\%.\par

\section{Discussion}
We discovered a QPO in 4U~1746--37 with a frequency of $\sim$1 Hz, and
an rms amplitude of $\sim$7.5\%. The QPO was also observed during type
I X-ray bursts where the count rate was a factor of 10 higher, with
similar fractional rms amplitudes. The QPO in 4U~1746--37 was only
observed when the source intensity was low. The frequency of the QPO,
the weak energy dependence of its rms amplitude, and the ratio Q
between the QPO frequency and its FWHM are similar to the frequencies,
the rms amplitude energy dependence, and Q-values (3--4) of the QPOs
discovered recently in the persistent emission, dips, and during the
type I X-ray bursts in the dippers and LMXBs 4U~1323--62 (Jonker et
al. 1999a) and in EXO~0748--676 at low intensity (Homan et
al. 1999). The similarities in QPO properties and the fact that
4U~1746--37 is a dipper suggest that the $\sim$1 Hz QPO is a general
property of high-inclination LMXBs at low intensities, and has a
common origin in these systems. \par

We found for the first time that 4U~1746--37 showed spectral
characteristics of an atoll source; the timing properties of
observations 4--9 were consistent with banana branch behavior reported
for atoll sources (van der Klis 1995). Although no band limited noise
was detected, upper limits (1--512 Hz) in the low intensity (hard)
state were 20\%--28\%, whereas they were 2.5\%--5.5\% when the source
was on the apparent banana branch.  So, with respect to the band
limited noise the timing behavior of the source is consistent with
that of an atoll source. Also the fact that on the banana branch the
rms amplitude of the power law noise component increased as the source
moved up in count rate is consistent with the behavior of the VLFN on
the banana branch of an atoll source. However, the high (5.9\%--7.1\%)
rms amplitude and low power law index (0.5) of the power law noise
component when the source is at low intensities is not consistent with
island-state VLFN.  We note that this could be due to effects
observable only at high inclinations, similar to those which cause the
dips and the $\sim$1 Hz QPO. In EXO~0748--676 strong power law noise
(6\%--11\%) with indices 0.5--0.8 was also present. This noise had no
or weak dependence on photon energy, just as the $\sim$1 Hz QPOs
(Homan et al. 1999), suggesting a similar origin for both
phenomena. In the one observation of EXO~0748--676 where no QPO was
found the rms amplitude of the VLFN was lower ($\sim$4\%) and the
power law index was $\sim$1. The spectral changes we report in
4U~1746--37 related to changes in the position along the atoll,
probably reflect the changing spectral characteristics reported before
for this source (eg. see Parmar et al. 1999).\par

Homan et al. (1999) found that the QPO in EXO~0748--676 was present in
all observations except one, where the persistent emission count rate
was $\sim$2.5 times higher than in the other observations. The fact
that the QPO is not detected at higher source count rates (and
inferred mass accretion rates) may be due to changes in the accretion
geometry with $\dot{M}$. We expect that if we observe 4U~1323--62 at
higher $\dot{M}$, the QPO may disappear as well. In 4U~1746--37 the
QPO also disappeared when the count rate increased. The rms amplitude
dropped by at least a factor of five when the source moved from the
low intensity state to the banana branch. This can be accounted for if
the size of the central source increased by a factor of 2--3 in
radius, decreasing the modulated fraction of the X-rays below our
treshold of 1\%--2\%. If this is the explanation the dip fraction
should decrease, making the dips more shallow. Since in 4U~1746--37
the dips were shallow and the source count rates were low, we were not
able to check this prediction for this source. Another possibility is
that the structure in or on the disk responsible for the $\sim$1 Hz
QPO disappeared or decreased in size, or the optical depth of the gray
medium changed.\par

We found that the burst e-folding time is shorter ($\sim$10~s) when
the source is in the low intensity state than when the source is on
the higher intensity banana branch ($\sim$30~s). This is opposite to
what was found by van Paradijs, Penninx, \& Lewin (1988) for several
sources. Van der Klis et al. (1990) found for 4U~1636--53 that the
burst duration was longer ($>20$~s) when the source was in the island
part of the CD than when the source was in the banana branch
($<10$~s). Either the state where we observed the short burst
e-folding times is not the product of an island state and high
inclination effects as we propose, but a state at a higher mass
accretion rate than the banana branch mass accretion rate, or another
parameter rather than $\dot{M}$ is also affecting the burst
duration.\par

The frequency of the QPO was 0.35 Hz higher during the type I X-ray
bursts than during the persistent emission of observation 1. Similar
increases in the QPO frequency during bursts were also observed in
EXO~0748--676 by Homan et al. (1999).  \par

\acknowledgments This work was supported by the Netherlands
Organization for Scientific Research (NWO) under contract number
614-51-002, and by NWO Spinoza grant 08-0 to E.P.J.van den
Heuvel. This research has made use of data obtained through the High
Energy Astrophysics Science Archive Research Center Online Service,
provided by the NASA/Goddard Space Flight Center. MM is a fellow of
the Consejo Nacional de Investigaciones Cient\'{\i}ficas y T\'ecnicas
de la Rep\'ublica Argentina.

\end{document}